\documentclass[aps,prl,twocolumn,superscriptaddress,groupedaddress]{revtex4}  
\usepackage{graphicx}  
\usepackage{dcolumn}   
\usepackage{bm}        
\usepackage{amssymb}   

\def\nn{\nonumber}

\newcommand{\be}{\begin{equation}}
\newcommand{\ee}{\end{equation}}
\def\ba{\begin{array}}
\def\ea{\end{array}}

\def\bo1{ \left | B^0 (p^+) \right \rangle}

\newcommand{\bea}{\begin{eqnarray}}
\newcommand{\eea}{\end{eqnarray}}

\begin{document}

\widetext
\title{The Holographic F Theorem }
\author{Marika Taylor and William Woodhead \\ STAG and Mathematical Sciences, University of Southampton, Highfield SO17 1BJ, UK}

\begin{abstract}
The F theorem states that, for a unitary three dimensional quantum field theory, the F quantity defined in terms of the 
partition function on a three sphere is positive, stationary at fixed point and decreases monotonically along a renormalization group flow. 
We construct holographic renormalization group flows corresponding to relevant deformations of three-dimensional conformal field theories on 
spheres, working to quadratic order in the source. For these renormalization group flows, the F quantity at the IR fixed point is always less than F at the UV fixed point, but F increases 
along the RG flow for deformations by operators of dimension $3/2 < \Delta < 5/2$. Therefore the strongest version of the F theorem is in general violated. 
 
\end{abstract}

\maketitle

In a three-dimensional quantum field theory, the F quantity is defined in terms of the renormalized partition function of the theory on a three-sphere $Z_{S^3}$ as 
\be
F = - \ln Z_{S^3}; \label{Fquanty}
\ee
$F$ gives the free energy on the three-sphere. 
The conjectured F-theorem \cite{Jafferis:2011zi,Klebanov:2011gs} states that $F$ is positive in a unitary quantum field theory; $F$ is stationary at a fixed point; $F_{UV} \ge F_{IR}$ for UV and IR fixed points and $F$ decreases monotonically along an RG flow.  Evidence in favour of the F theorem has been presented in a number of works. In \cite{Jafferis:2011zi} it was shown in a number of ${\cal N} = 2$ theories that $F_{IR} < F_{UV}$; examples included holographic theories described by $AdS_4 \times Y_7$ M theory solutions in which the partition function is \cite{Herzog:2010hf}
\be
F = N^{\frac{3}{2}} \sqrt{\frac{2 \pi^6}{27 {\rm Vol}(Y_7)}} \label{scaling}
\ee
where ${\rm Vol}(Y_7)$ is the volume of the Sasaki-Einstein manifold $Y_7$ and $N$ is the number of colors in the dual theory. 

Many subsequent papers have provided additional evidence that $F_{IR} < F_{UV}$ in holographic and field theory models. For example, \cite{Klebanov:2011gs} considered relevant double trace deformations: given an operator $\Phi$ in a CFT of dimension $\Delta_{-}$ such that $1/2 \le \Delta_- \le 3/2$,  deforming the CFT by $\Phi^2$ causes an RG flow to an IR fixed point where $\Phi$ has dimension $\Delta_{+} = 3 - \Delta_{-}$. 

The evidence for stationarity and monotonic decrease of the F quantity along an RG flow is somewhat weaker. Arguments for stationarity are based on the fact that F is extremised with respect to the R charges of an IR CFT \cite{Jafferis:2010un} and (holographically) with respect to the parameters of the Sasaki-Einstein manifold $Y_7$ \cite{Martelli:2011qj,Jafferis:2011zi}. For monotonic decrease, it was shown in \cite{Klebanov:2011gs} that the free energy decreases monotonically along weakly relevant flows, while \cite{Gulotta:2011si} argued that the volume of the compact manifold should increase monotonically along an RG flow in holographic examples, implying monotonic decrease of F. In \cite{Freedman:2013ryh} it was shown that F decreases along certain supersymmetric RG flows of deformations of the ABJM theory. 

In a conformal field theory, the partition function on the three sphere is related to the (finite terms) in the entanglement entropy for a disk region in flat space by the Casini-Huerta-Myers map \cite{Casini:2011kv}. If the finite contribution to the entanglement entropy of a disk region in the ground state of the CFT is
\be
S = - 2 \pi {\cal F}, 
\ee
then ${\cal F}$ corresponds precisely to the F quantity, i.e. ${\cal F}$ is conjectured to be positive and to decrease monotonically along an RG flow. The F theorem has hence also been explored using entanglement entropy, see for example \cite{Klebanov:2011td,Casini:2012ei,Klebanov:2012yf}. Ambiguities in defining the finite contributions can be dealt with by working with the UV finite mutual information \cite{Casini:2015woa} or by using renormalized entanglement entropy \cite{Liu:2012eea}. There is however evidence that the renormalized entanglement entropy thus defined is not stationary at a fixed point \cite{Klebanov:2012va}. 

In this paper we show that the F quantity does not decrease for holographic RG flows associated with deformations by single trace operators of dimension  $d/2 < \Delta_{+} < d$. Therefore the strong version of the F theorem, decrease of F under all relevant deformations, is false.  Note that in all our examples a weaker version of the F theorem, $F_{UV} \ge F_{IR} > 0$, is still satisfied. 


We begin by discussing holographic realisations of RG flows on curved manifolds. We work in Euclidean signature with a bulk action
\be
I_E = - \frac{1}{16 \pi G_4} \int d^4 x \sqrt{g} \left ( R - \frac{1}{2} (\partial \phi)^2 + V(\phi) \right ),
\ee
where $G_4$ is the Newton constant, which in a top-down holographic model is related to the number of colors as $1/G_4 \sim N^{3/2}$, as in (\ref{scaling}). 
We consider solutions of the equations of motion such that 
\be
ds^2 = dw^2 + e^{2 {\cal A}(w)} d s_{\Omega_3}^2
\ee
where $\Omega_3$ is a homogeneous space with Ricci scalar ${\cal R}$ and the scalar field $\phi$ depends only on the radial coordinate $w$. We will 
be interested in the case of a unit radius three sphere for which ${\cal R} = 6$. The equations of motion are then given by:
\bea
\ddot{\phi} + 3 \dot{\cal A} \dot{\phi} &=& - V'(\phi); \\
- \frac{ {\cal R}}{6} e^{-2 {\cal A}} - \frac{1}{4} (\dot{\phi})^2 &=& \ddot{\cal A}. \nn
\eea
These equations reduce to the case of flat domain walls when ${\cal R} = 0$.


We work perturbatively in the scalar field and assume that the potential has the following analytic expansion in $\phi$ around an AdS background:
\be
V(\phi) = 6 - \frac{1}{2} M^2 \phi^2 + \cdots
\ee
In what follows we solve the field equations to quadratic order in $\phi$, taking into account the backreaction onto the metric to this order. In anti-de Sitter the warp factor is:
\be
{\cal A}(w) \equiv {\cal A}_0(w) = \log ( \sinh (w)).
\ee
Working to quadratic order in the scalar field, the change in the warp factor is quadratic in the scalar field, and therefore to the order required the scalar field equation is
that in AdS, i.e.
\be
\ddot{\phi} + 3  \coth w \dot{\phi} = M^2 \phi.
\ee 
This equation can be solved exactly (see below) and asymptotically near the conformal boundary. The latter can be expressed as 
\bea
\phi &=& e^{ (\Delta_+ - 3) w} \phi_{(0)} + e^{(\Delta_+ - 5)w} \phi_{(2)} + \cdots  \label{f-exp2} \\
&& + e^{-\Delta_+ w} \tilde{\phi}_{(0)} + e^{ (-\Delta_+ - 2) w} \tilde{\phi}_{(2)} + \cdots  \nn
\eea 
where 
\be
\phi_{(2)} = \frac{3(3 - \Delta_+)}{(5 - 2\Delta_+)} \phi_{(0)} \qquad \tilde{\phi}_{(2)} = \frac{3 \Delta_+ \phi_{(0)}}{(2 \Delta_+ -1)}. 
\ee 
Here we implicitly assume that $\Delta_+$ is neither $3/2$ nor $5/2$, since in these cases terms proportional to $w$ arise in the expansion. (These cases can be straightforwardly analysed but we do not include details in what follows.)

One can then use the other equation of motion to solve for the warp factor up to quadratic order in the scalar field. Letting
\be
{\cal A} = {\cal A}_{0} + a
\ee
then 
\be
\ddot{a} - \frac{2}{\sinh^2 w} a = - \frac{1}{4} \dot{\phi}^2. 
\ee
The onshell action is divergent for asymptotically locally AdS solutions, but the divergences may be removed by using the asymptotic solutions of the field equations to regulate 
the bulk action and adding appropriate covariant counterterms i.e. the renormalized action  \cite{DeHaro2001}.
\be
I_{\rm ren} = I^{\rm}_{E} + I_{\rm ct} \label{iren}
\ee
is finite.  The AdS/CFT dictionary implies that the F quantity is calculated from the renormalized action in the limit in which the dual theory is well-described by supergravity. 

Working to quadratic order in the scalar field $\phi$ the required counterterms to render the action finite are \cite{DeHaro2001}
\bea
I_{\rm ct} &=& \frac{1}{8 \pi G_4} \int d^3 x \sqrt{h} \left ( - K + 2 + \frac{1}{2} {\cal R}_h  \right . \label{bct1} \\
&& \qquad  \qquad \left . + \frac{1}{4} (3 - \Delta_+) \phi^2 +  \frac{(\Delta_+ - 3)}{16 (2 \Delta_+ - 5)} {\cal R}_h \phi^2 \right ) \nonumber
\eea
where ${\cal R}_h$ is the Ricci scalar for the boundary metric $h$.  We define $\Delta_+$ in terms of the mass as
\be
\Delta_+ = \frac{3}{2} + \frac{1}{2} \sqrt{9 + 4 M^2}
\ee
and we assume that $\frac{3}{2} \le \Delta_+ \le 3$. For $\Delta_{+} \ge 5/2$, $\Delta_{+}$ is the dimension of the operator dual to the scalar field of mass $M^2$. In the mass range
\be
- \frac{9}{4} \le M^2 \le - \frac{5}{4}
\ee
two quantizations are possible \cite{Klebanov:1999tb}; we will discuss this situation below. In (\ref{bct1}) we do not include counterterms which depend on derivatives
of the scalar field (see \cite{DeHaro2001}), since the scalar fields under consideration are homogeneous. 

Note that the last counterterm in (\ref{bct1}) is only required for $\Delta_+ > 5/2$. 
The corresponding divergence becomes logarithmic at $\Delta_+ = 5/2$ and in this case the value of the renormalised action can be adjusted by finite counterterms, so the F quantity is inherently scheme dependent. Correspondingly F is also scheme dependent for the $\Delta_-$ quantization of the same mass, i.e. $\Delta_- = 1/2$. No finite counterterms arise for other values of $\Delta_+$ in the range of interest, although working to cubic order in the scalar field finite counterterms would arise at integral values of $\Delta_+$; these can be fixed by requiring supersymmetry \cite{Freedman:2013ryh}.

In the mass range $-9/4 \le M^2 \le - 5/4$, two quantizations are possible:
\be
\Delta_{\pm} = \frac{3}{2} \pm \sqrt{9 + 4 M^2}
\ee
with $1/2 \le \Delta_- \le 3/2$ and $3/2 \le \Delta_+ \le 5/2$. 
As discussed in  \cite{Papadimitriou:2006dr}, the evaluation of the renormalized action by adding covariant counterterms is not affected by whether the dual operator has dimension $\Delta_{+}$ or $\Delta_{-}$. The difference arises in the identification of the functional that generates correlation functions for the dual operator. 
For the ${\Delta_+}$ quantization, the coefficient $\phi_{(0)}$ in (\ref{f-exp2}) acts as the source for the dual operator. The renormalized action (\ref{iren}) is a functional of this coefficient and acts as the generating functional for the dual operator. 

For the ${ \Delta_-}$ quantization, the renormalized action (\ref{iren}) is still a functional of the coefficient $\phi_{(0)}$ in (\ref{f-exp2}) but this coefficient is not the operator source. As discussed in \cite{Papadimitriou:2006dr}, following \cite{Klebanov:1999tb,Witten:2001ua}, the correct generating functional is obtained by a Legendre transformation. Let us define the Legendre transformation as 
\be
\tilde{I} [\phi_{(0)}, {\psi}_{(0)} ] = I_{\rm ren} [\phi_{(0)}] + \int d^3 x \sqrt{g_{(0)}} \phi_{(0)} {\psi}_{(0)}
\ee
where $g_{(0)}$ is the boundary metric. Then extremising gives
\be
\tilde{I}_{\rm ren}[{\psi}_{(0)}] = \tilde{I}[\phi^{\ast}_{(0)}({\psi}_{(0)}), {\psi}_{(0)}]  
\ee
where 
\be
\left . \frac{\delta I_{\rm ren} [\phi_{(0)}]}{\delta \phi_{(0)}} \right |_{ \phi^{\ast}_{(0)}} + {\psi}_{(0)} = 0
\ee
defines $\phi^{\ast}_{(0)}({\psi}_{(0)})$. Here $\tilde{I}_{\rm ren}[{\psi}_{(0)}]$ is identified as the renormalized generating functional of correlation functions of the operator of dimension $\Delta_-$.

In the case at hand, we work perturbatively in the scalar field and thus the onshell renormalized action necessarily has the form
\be
I_{\rm ren} [\phi_{(0)}] = (I_0 + I_2 \phi_{(0)}^2 + \cdots ), \label{fren1}
\ee
where $I_0$ and $I_2$ are numerical coefficients. (Recall that $\phi_{(0)}$ is homogeneous and therefore does not depend on the sphere coordinates.)
The Legendre transformed action is then given by 
\be
\tilde{I} [\phi_{(0)}, \Psi_{(0)} ] = (I_0 + I_2 \phi_{(0)}^2 + \cdots) +  \phi_{(0)}{\Psi}_{(0)},
\ee
where we denote
\be
{\Psi}_{(0)} \equiv  \int d^3 x \sqrt{g_{(0)}}  {\psi}_{(0)}.
\ee
Extremising, we obtain 
\be
2 I_2 \phi_{(0)} + {\Psi}_{(0)} = 0
\ee
and hence 
\be
\tilde{I}_{\rm ren}[{\Psi}_{(0)}]  = I_0 - I_2 \phi_{(0)}^2 + \cdots = I_0 - \frac{1}{4 I_2} {\Psi}_{(0)}^2 + \cdots \label{fren2}
\ee
In the $\Delta_{+}$ quantization, $\phi_{(0)}$ acts as the source for the dual operator and therefore (\ref{fren1}) gives the free energy to quadratic order in the source. 
In the $\Delta_{-}$ quantization, ${\Psi}_{(0)}$ acts as the source for the dual operator and (\ref{fren2}) gives the free energy to quadratic order in the source. The quadratic terms have different signs in the two quantizations: if $I_2 > 0$ the free energy on the $S^3$ increases for deformations by the $\Delta_{+}$ quantization operator and decreases for deformations by the $\Delta_{-}$ quantization operator, and vice versa. For $M^2 = - 9/4$, the two quantizations coincide; note however that one needs to treat this case separately, as the above formulae degenerate.


Having determined the renormalized free energy functional we now consider exact regular solutions of the field equations, to quadratic order in the scalar field. To the required order we can solve the scalar field equation in the anti-de Sitter background. The scalar field solution may be found analytically: 
\bea
&& \phi = \frac{\phi_{(0)} \Gamma\left(\frac{5}{2}-\Delta_+\right)}{2 \sqrt{2} \Gamma(3 - \Delta_+)} {(U-1)}^{(3 - \Delta_+)/2} {(U+1)}^{\Delta_+/2} \\
&& \left[ \Gamma(3-\Delta_+) {_2 \tilde{F}_1}\left(- \frac{1}{2}, \frac{3}{2}; \frac{5}{2} - \Delta_+; \frac{1}{2}(1-U)\right) \right. \nonumber \\
&&-\left. \left(\frac{U-1}{U+1}\right)^{\Delta_+-\frac{3}{2}} \Gamma(\Delta_+) {_2 \tilde{F}_1}\left(- \frac{1}{2}, \frac{3}{2}; \Delta_+ - \frac{1}{2}; \frac{1}{2}(1-U)\right) \right] \nonumber \\
&& U = \frac{1}{u} = \coth{w} \nonumber
\eea
where we have imposed regularity throughout the bulk and chosen the overall normalisation of the solution to agree with the definition of $\phi_{(0)}$ given in (\ref{f-exp2}). 

\begin{figure}[tbp]
  \centering
  \includegraphics[scale=0.6]{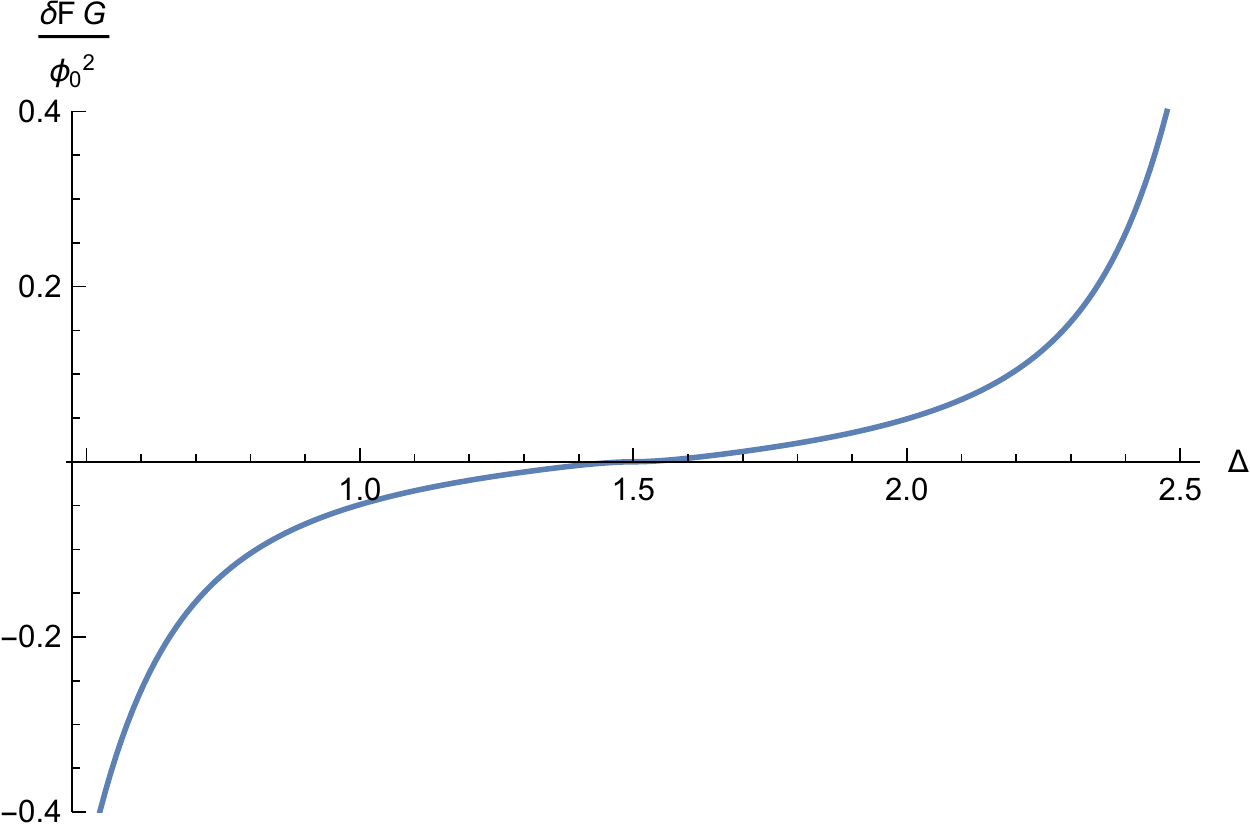}
  \caption{The change in the renormalized free energy for $\frac{1}{2} < \Delta < \frac{5}{2}$. }\label{fig:free_energy_lower_range}
\end{figure}
\begin{figure}[tbp]
  \centering
  \includegraphics[scale=0.6]{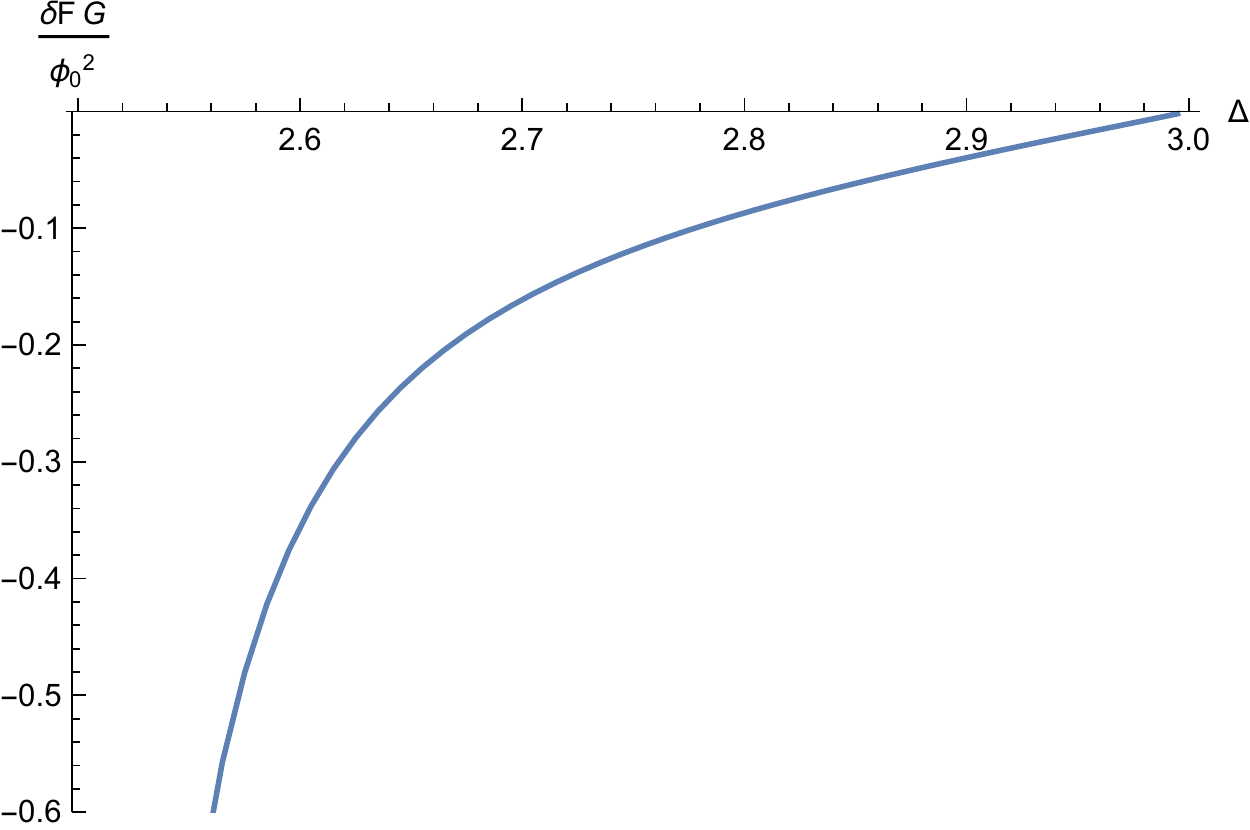}
  \caption{The change in the renormalized free energy for $\frac{5}{2} < \Delta < 3$. }\label{fig:free_energy_upper_range}
\end{figure}

As $w \rightarrow 0$, 
\be
\phi \rightarrow \phi_{\rm IR} \left ( 1 + \frac{1}{5} M^2 w^2 + \cdots \right )
\ee
where
\be
\phi_{\rm IR} =\frac{1}{4 \sqrt{\pi}} \Gamma(\frac{5}{2} - \Delta_+) \Gamma(\Delta_+) \cos(\pi \Delta_+) \phi_{(0)}
\ee
and consequently the change in the warp factor behaves as 
\be
a \rightarrow - \frac{M^4}{250} \phi_{\rm IR}^2 w^4 + \cdots
\ee
as $w \rightarrow 0$, i.e. the geometry in the deep interior approaches (Euclidean) $AdS_4$ in spherical coordinates, and therefore the RG flow ends on an IR fixed point. For such an IR fixed point, the free energy as computed from the renormalized action is
\be
F_{IR} = \frac{\pi}{2 G_4} \left (1 - \frac{1}{12} M^2 \phi_{\rm IR}^2 \right )
\ee
which clearly satisfies $F_{IR} < F_{UV}$: the free energy for the UV fixed point is given by $\phi_{\rm IR} = 0$. 

To calculate the free energy for the RG flow, we need to solve numerically for the warp factor and thus for the renormalized onshell action. 
To carry out the numerics we work with a compactified radial coordinate $u = \tanh w$ for all the calculations. Plotted in Figures~\ref{fig:free_energy_lower_range}--\ref{fig:free_energy_upper_range} is the change in the free energy normalised by the scalar source and the Newton constant: we define $\delta F$ as  
\be
\delta F = F(\phi_{(0)}) - F(0) \equiv F(\phi_{(0)}) -  \frac{\pi}{2 G_4},
\ee
where implicitly we use the appropriate source for $\Delta < 3/2$. 

The change in the free energy vanishes to quadratic order in the source for $\Delta = 3/2$ and
for exactly marginal operators. 
The change in the free energy is however positive for $3/2 < \Delta < 5/2$, with the corresponding change in the free energy for $1/2 < \Delta < 3/2$ therefore being negative, in agreement with the arguments above. 

We should note that a related sign change at $\Delta = 3/2$ was found in \cite{Berenstein:2014cia}. This paper calculated one point functions of the deformation operator under relevant deformations, both holographically and using conformal perturbation theory. The sign of this one point function, which is related to the sign of $\delta F$ found above, indeed changes at $\Delta = d/2$.  

Thus, to summarise, working quadratically in the operator source, deformations by operators of dimensions $3/2 < \Delta < 5/2$ lead to increases in the F quantity, although the corresponding IR fixed points still satisfy $F_{IR} < F_{UV}$. Note that we have worked only to quadratic order and changes in the F quantity to higher order in the source would depend on the interactions in the theory. 

The F theorem would be satisfied in a holographic theory which contains no operators of dimensions $3/2 < \Delta < 5/2$, but generically such operators do exist.
In particular, it is well-known that in four-dimensional ${\cal N} = 8$ gauged supergravity there are 35 $\Delta_-=1$ scalar operators and 35 $\Delta_+ = 2$ pseudoscalars corresponding to the seventy scalars with $M^2 = -2$, i.e. both quantizations arise \cite{Hawking:1983mx}. (The pseudoscalar nature of the $\Delta_{+}$ quantization does not affect the arguments given here.) However, for the supersymmetric RG flows in consistent truncations of ${\cal N} = 8$ analysed in \cite{Freedman:2013ryh} the F quantity does decrease: in this setup supersymmetry does not allow a single real scalar (in the Euclidean) corresponding to a $\Delta_+ = 2$ operator to be switched on. A complete proof of the F theorem would effectively restrict the allowed holographic theories, i.e. it would throw theories such as those considered here into the swampland. 

Now let us return to the relationship between the F quantity and the entanglement entropy of a disk entangling region. One can use holographic renormalization techniques to 
define renormalized entanglement entropy \cite{Taylor:2016c}. The renormalized entanglement entropy of disk regions in theories deformed by relevant operators agrees with the behaviour of the F quantity found above: F increases for RG flows by operators of dimension $3/2 < \Delta < 5/2$ (see also \cite{Nishioka:2014kpa}).

Our results do not contradict \cite{Martelli:2011qj,Jafferis:2011zi}: these works showed that F is extremal within the parameter spaces of putative conformal field theories. In the holographic setups, an $AdS_4$ factor is assumed and the volume of the compactifying Sasaki-Einstein is extremised. This analysis does not imply that $F$ is decreased under relevant deformations which change the geometry away from $AdS_4$. 
The scalar field $\phi$ and the change in the warp factor $a$ do not decrease monotonically along the flow but this does not in itself contradict the arguments of \cite{Gulotta:2011si}. From a top-down perspective scalar fields in four-dimensional gauged supergravity theories arise not just from breathing modes of the compact manifold, but also from the four-form flux in eleven dimensions.

 In two dimensions one defines the Zamolodchikov c-function $c(g_i,\mu)$ in terms of the coupling constants $g_i$ and the energy scale $\mu$. Here implicitly we have defined F as a renormalized quantity, dependent on UV data for coupling constants of the relevant operators. It would be interesting to explore whether one could sharply define an F function with explicit dependence on the energy (i.e. radial) scale holographically.  One natural way to do this would be to rewrite the source $\phi_{(0)}$ in terms of the bulk scalar field $\phi(w)$, and interpret $w$ as the energy scale. 
 
 Finally, there has been considerable recent interest in how much supersymmetry is required to determine uniquely partition functions on even-dimensional spheres \cite{Gomis:2015yaa}. In three dimensions, an analogous question arises at quadratic order in the source for conformal field theories deformed by operators of dimensions $1/2$ and $5/2$. 


\section*{Acknowledgments}

This work was supported by the Science and Technology Facilities Council. 
This project has received funding from the European UnionÕs Horizon 2020
research and innovation programme under the Marie Sklodowska-Curie grant
agreement No 674896.

\end{document}